\documentclass[conference]{IEEEtran}

\usepackage{cite}
\usepackage{array} 

\usepackage{amsmath,amssymb,amsfonts}
\usepackage{algorithmic}
\usepackage{tabularx}
\usepackage{graphicx}
\usepackage{textcomp}
\usepackage{xcolor}
\usepackage{makecell}
\usepackage{booktabs} 
\usepackage{multirow} 
\usepackage{url}      
\usepackage{adjustbox} 
\usepackage{algorithm}

\usepackage[bookmarks=false, colorlinks=true, urlcolor=black, linkcolor=black, citecolor=black]{hyperref}

\def\BibTeX{{\rm B\kern-.05em{\sc i\kern-.025em b}\kern-.08em
    T\kern-.1667em\lower.7ex\hbox{E}\kern-.125emX}}

\newcommand{\stixtag}[1]{\texttt{#1}}
\begin{document}

\title{CyberNER: A Harmonized STIX Corpus for Cybersecurity Named Entity Recognition}


\author{
    \IEEEauthorblockN{
        Yasir ECH-CHAMMAKHY\IEEEauthorrefmark{1}\IEEEauthorrefmark{2},
        Anas Motii\IEEEauthorrefmark{1},
        Anass Rabii\IEEEauthorrefmark{2},
        Oussama Azrara\IEEEauthorrefmark{3},
        Jaafar Chbili\IEEEauthorrefmark{3}
    }
    \IEEEauthorblockA{
        \IEEEauthorrefmark{1}College of Computing, Mohammed VI Polytechnic University (UM6P), Ben Guerir, Morocco \\
        \IEEEauthorrefmark{2}Deloitte Morocco Cyber Center, Casablanca, Morocco \\
        \IEEEauthorrefmark{3}Deloitte Conseil, Paris, France
    }
    \IEEEauthorblockA{
        Emails: \IEEEauthorrefmark{1}\{Yasir.ECH-CHAMMAKHY, Anas.MOTII\}@um6p.ma;
        \IEEEauthorrefmark{2}arabii@deloitte.fr;
        \IEEEauthorrefmark{3}\{oazrara, jchbili\}@deloitte.fr
    }
}

\maketitle

\begin{abstract}
Extracting structured intelligence via Named Entity Recognition (NER) is critical for cybersecurity, but the proliferation of datasets with incompatible annotation schemas hinders the development of comprehensive models. While combining these resources is desirable, we empirically demonstrate that naively concatenating them results in a noisy label space that severely degrades model performance. To overcome this critical limitation, we introduce \textbf{CyberNER}, a large-scale, unified corpus created by systematically harmonizing four prominent datasets (CyNER, DNRTI, APTNER, and Attacker) onto the STIX 2.1 standard. Our principled methodology resolves semantic ambiguities and consolidates over 50 disparate source tags into 21 coherent entity types. Our experiments show that models trained on CyberNER achieve a substantial performance gain, with a relative F1-score improvement of approximately 30\% over the naive concatenation baseline. By publicly releasing the CyberNER corpus, we provide a crucial, standardized benchmark that enables the creation and rigorous comparison of more robust and generalizable entity extraction models for the cybersecurity domain.
\end{abstract}


\section{Introduction}
\label{sec:introduction}

Named Entity Recognition (NER), the task of identifying and classifying predefined entities within unstructured text, is a key technology for automating the extraction of structured knowledge from the vast amounts of textual data generated in the cybersecurity domain. Accurate NER systems are crucial for various downstream Cyber Threat Intelligence (CTI) applications, including indicator extraction, threat actor profiling, vulnerability analysis, knowledge graph construction, and automated report generation~\cite{arazzi2023nlp, evangelatos2021ner}. By recognizing key entities like malware families, CVE identifiers, threat actor groups, attack patterns, tools, and Indicators of Compromise (IOCs), NER transforms raw text into machine-processable intelligence.

In recent years, the research community has produced several valuable annotated datasets specifically for training and evaluating cybersecurity NER models. Resources such as CyNER~\cite{alam2022cyner}, DNRTI~\cite{wang2020dnrti}, STUCCO~\cite{bridges2014automatic}, AttackER~\cite{deka2024attacker}, and APTNER~\cite{wang2022aptner}, among others, provide labeled examples covering diverse aspects of the cyber threat landscape. While the availability of these datasets represents significant progress, a major obstacle limits their collective utility: significant heterogeneity in annotation schemas and entity definitions.

Each dataset was typically created with specific goals or CTI perspectives in mind, resulting in different sets of entity types, varying levels of granularity, and sometimes conflicting definitions for similar concepts. For example, threat actors might be labeled with a general \texttt{Threat\_Actor} tag, a specific group type like \texttt{APT} or \texttt{HackOrg}, or even a more generic \texttt{IDTY} (Identity) tag depending on the dataset's focus. Similarly, organizational entities might be tagged as \texttt{Organization}, \texttt{Org}, \texttt{IDTY}, or even distinguished as \texttt{VICTIM\_IDENTITY} versus \texttt{GENERAL\_IDENTITY}. This lack of standardization makes it highly difficult to directly compare models trained on different datasets, prevents the straightforward combination of datasets to train more comprehensive NER models, and complicates efforts to build CTI tools that consume output from different NER systems. Naively concatenating these datasets leads to inconsistent labels for the same underlying entity, confusing models and degrading performance~\cite{llorca2023meta}.

To address this critical gap and promote progress in cybersecurity NER, we introduce CyberNER, a large-scale, harmonized corpus developed by systematically mapping entities from multiple prominent cybersecurity NER datasets onto a single, consistent taxonomy. Our unified taxonomy is based on the widely adopted STIX 2.1 standard~\cite{stix_intro}, chosen for its industry relevance, structured vocabulary for CTI objects and concepts, and potential to facilitate interoperability. This paper details the rigorous methodology employed for schema harmonization, including the analysis of source dataset schemas, the development of explicit mapping rules, and the systematic resolution of ambiguities and conflicts encountered when aligning disparate tags with the STIX framework.

The main contributions of this paper are threefold:
\begin{itemize}
    \item We introduce and publicly release CyberNER\footnote{The CyberNER dataset, along with code for data harmonization and model benchmarking, are publicly available at \url{https://github.com/yasirech-chammakhy/CyberNER}.}, a large-scale, harmonized NER corpus for the cybersecurity domain. It integrates annotations from four diverse datasets onto a unified STIX 2.1-based taxonomy comprising 21 distinct entity types and contains approximately 610k tokens.
    \item We conduct an empirical study that quantifies the performance degradation resulting from naive dataset concatenation, establishing a crucial baseline that highlights the necessity of schema harmonization.
    \item We provide an extensive benchmark of state-of-the-art models on CyberNER and show that our harmonization methodology yields a substantial performance increase, with a relative F1-score improvement of approximately 30\% over the naive baseline. This validates the efficacy of our approach and sets a new, robust benchmark for future research.
\end{itemize}

By providing this unified, STIX-aligned benchmark dataset, CyberNER aims to facilitate the development and rigorous evaluation of more robust and generalizable NER models for the cybersecurity domain, ultimately contributing to more effective automated CTI processing.

The remainder of this paper is structured as follows: Section~\ref{sec:related_work} reviews prior research in the field. Section~\ref{sec:harmonization_methodology} details our schema analysis and harmonization process. Section~\ref{sec:dataset_description} describes the resulting CyberNER corpus. Section~\ref{sec:experiments} presents our comparative experiments and results. Finally, Section~\ref{sec:discussion} discusses our findings and limitations, and Section~\ref{sec:conclusion} concludes the paper.

\section{Related Work}
\label{sec:related_work}

The evolution of NER models in cybersecurity has mirrored broader advancements in NLP, moving from conventional methods to powerful deep learning architectures~\cite{Elouardi2025OptiFLIDS}. Early approaches often combined rule-based systems with feature-engineered statistical models. Rule-based strategies utilized gazetteers (curated entity lists) and regular expressions to capture well-structured entities like CVEs, IP addresses, and file hashes, a technique effectively employed by Hanks et al.~\cite{hanks2022recognizing} and Alam et al.~\cite{alam2022cyner}. Alongside these, statistical models like Hidden Markov Models (HMMs) and Support Vector Machines (SVMs) were applied, but Conditional Random Fields (CRFs) became the standard for sequence labeling due to their ability to model dependencies between adjacent tags~\cite{lafferty2001conditional}.

The paradigm shifted with the advent of deep learning, which circumvents the need for laborious feature engineering. Architectures combining Bidirectional Long Short-Term Memory (BiLSTM) networks with a CRF layer (BiLSTM-CRF) became a de facto standard, effectively capturing long-range contextual information within sentences~\cite{huang2015bidirectional}. Researchers quickly adapted these models for cybersecurity, often enhancing them with character-level features~\cite{kim2015characteraware} and domain-specific word embeddings (e.g., Word2Vec, GloVe) trained on security corpora to improve performance on specialized datasets~\cite{guo2021key}.

More recently, Transformer-based architectures~\cite{vaswani2017attention} have surpassed sequential models. The remarkable success of BERT (Bidirectional Encoder Representations from Transformers)~\cite{devlin2019bert, Amadou2024EUREKHA} spurred a new wave of research focused on fine-tuning pre-trained language models for CTI tasks. This led to the development of several powerful, domain-specific models pre-trained on massive cybersecurity corpora to better understand the domain's unique terminology. Notable examples include SecureBERT~\cite{aghaei2022securebert}, trained on over 2.2 million security documents; CySecBERT~\cite{fiorini2023cysecbert}, adapted from cybersecurity-related social media and reports; and DarkBERT~\cite{jin2023darkbert}, uniquely trained on a large corpus of dark web data. Studies consistently show that combining these powerful transformer encoders with a CRF layer yields state-of-the-art results on various cybersecurity NER benchmarks~\cite{wang2022aptner, demirol2025novel}.

Despite these advanced models, performance remains constrained by a fragmented landscape of NER datasets, each with a unique, non-interoperable schema (e.g., CyNER~\cite{alam2022cyner}, DNRTI~\cite{wang2020dnrti}, APTNER~\cite{wang2022aptner}). A recent study by Schmidt et al.~\cite{jalocha2025label} highlighted this challenge, demonstrating that a coarse-grained, ad-hoc unification of these datasets results in poor cross-dataset generalization. Their work underscores the failure of simplistic merging. In contrast, our research takes a data-centric approach. We introduce a principled harmonization methodology using the rich, industry-standard STIX 2.1 taxonomy. We demonstrate that this ontology-driven approach, unlike naive unification, successfully creates a coherent corpus that yields substantial performance gains, addressing a key data-level bottleneck in the field.

\section{Schema Harmonization Methodology}
\label{sec:harmonization_methodology}

We systematically harmonized multiple heterogeneous datasets into the coherent, STIX-aligned CyberNER corpus. This involved analyzing source datasets, defining a target taxonomy based on STIX 2.1, developing explicit mapping rules, standardizing data formats, and performing quality control.

\subsection{Source Dataset Analysis}
\label{sec:source_analysis}

We began by analyzing the selected source datasets: CyNER~\cite{alam2022cyner}, DNRTI~\cite{wang2020dnrti}, APTNER~\cite{wang2022aptner}, and Attacker~\cite{deka2024attacker}. As summarized in Table~\ref{tab:dataset_stats}, this analysis revealed significant heterogeneity, preventing straightforward integration. Key challenges included:

\begin{itemize}
    \item \textbf{Varying Entity Sets and Granularity:} Datasets define incompatible entity types (e.g., 5 types in CyNER vs. 23 fine-grained types like \texttt{MD5} and \texttt{VULID} in APTNER) and introduce unique categories (e.g., \texttt{HackOrg} in DNRTI, \texttt{COURSE\_OF\_ACTION} in Attacker). This prevents direct merging due to differing scope and detail levels.

    \item \textbf{Inconsistent Labeling:} Similar concepts lack uniform tags. For instance, threat actors are labeled \texttt{APT}, \texttt{HackOrg}, or \texttt{THREAT\_ACTOR}, and malware/tools have varied names across datasets.

    \item \textbf{Differing Tag Formats:} Source datasets employ incompatible sequence tagging schemes, notably BIOES (APTNER) versus BIO (others), necessitating conversion to a standard format.

    \item \textbf{Formatting Issues:} Preprocessing identified noise such as malformed lines and occasional parsing errors (e.g., common words tagged as entities), which required rule-based cleaning.
\end{itemize}

These inconsistencies underscore the inadequacy of simple dataset concatenation and necessitate a structured harmonization process. This motivated our adoption of the STIX~2.1 standard as a unifying target taxonomy.

The selection of these four particular datasets was deliberate and strategic, aimed at creating a benchmark that encapsulates the diverse challenges present in real-world CTI analysis. Our selection criteria were threefold:  
\begin{enumerate}
    \item \textbf{Schema Diversity and Complexity:} The chosen datasets span a wide spectrum of annotation complexity, from the five broad entity types in CyNER to the 23 fine-grained, technical types in APTNER. This ensures that a model trained on the harmonized corpus is robust to varying levels of semantic granularity.
    \item \textbf{Domain Coverage:} The sources cover a range of cybersecurity sub-domains, including technical indicator reports (DNRTI), detailed APT campaign analysis (APTNER, Attacker), and broader cybersecurity news (CyNER). This textual diversity exposes models to different linguistic styles and reporting formats.
    \item \textbf{Community Adoption:} Each of these datasets is an established and widely used resource within the academic community, making our harmonized corpus a relevant and valuable successor for comparative research.
\end{enumerate}

By integrating these specific sources, we ensure that CyberNER is not just large but also a comprehensive and challenging testbed for developing truly generalizable NER models.

\begin{table}[htbp]
\centering
\caption{Summary Statistics and Schema Characteristics of Source Datasets (Pre-Harmonization).}
\label{tab:dataset_stats}
\footnotesize
\begin{tabular}{@{}lrrrr@{}}
\toprule
\textbf{Dataset} & \textbf{Tokens} & \textbf{Entities} & \textbf{Unique Types} & \textbf{Format} \\
\midrule
CyNER~\cite{alam2022cyner}     & 107k  & 7.6k  & 5  & BIO \\
DNRTI~\cite{wang2020dnrti}     & 176k  & 36.9k & 12 & BIO \\
APTNER~\cite{wang2022aptner}   & 260k  & 39.7k & 23 & BIOES \\
Attacker~\cite{deka2024attacker} & 67k   & 31.1k & 20 & BIO \\
\bottomrule
\end{tabular}
\vspace{-1em}
\end{table}

\subsection{Target Taxonomy Design using STIX 2.1}
\label{sec:target_taxonomy}

To resolve the schema heterogeneity identified in Section~\ref{sec:source_analysis}, we selected the Structured Threat Information eXpression (STIX) version 2.1~\cite{stix_intro} as the basis for our unified target taxonomy. STIX provides a standardized, community-approved language for representing Cyber Threat Intelligence (CTI), defining a structured vocabulary for concepts ranging from high-level STIX Domain Objects (SDOs) like \texttt{Threat-Actor} and \texttt{Malware} to specific STIX Cyber-observable Objects (SCOs) such as \texttt{IP-Address} and \texttt{File}.

Using STIX 2.1 offers crucial advantages for creating a robust and widely applicable cybersecurity NER dataset. Firstly, its standardized definitions provide semantic clarity, enabling us to resolve the naming conflicts and ambiguities found in the source datasets. Mapping diverse source tags (e.g., \texttt{APT}, \texttt{HackOrg}) to a single, clearly defined STIX object (\texttt{Threat-Actor}) ensures consistent labeling within CyberNER. Secondly, this STIX alignment promotes interoperability. Models trained on CyberNER can produce output readily usable by other CTI tools and platforms that adhere to the STIX standard, enhancing their practical utility.

Furthermore, the structured nature of STIX offers inherent adaptability and extensibility. While our primary goal is NER, STIX objects have defined relationships (e.g., malware \emph{used by} threat-actor). Using STIX-based labels provides a foundation for future work, such as extracting these relationships to build knowledge graphs~\cite{mouiche2025entity}. This alignment with an industry-standard framework also enhances the community relevance and usability of the CyberNER corpus.

Our final target taxonomy comprises 21 distinct entity labels, representing a curated subset of STIX 2.1 objects chosen for broad coverage of entities commonly found in CTI reporting and relevant to the source datasets.

\begin{table}[htbp] 
\centering
\caption{Illustrative Examples of Source Tag to STIX 2.1 Mapping}
\label{tab:mapping_examples_short} 
\footnotesize 
\setlength{\tabcolsep}{3pt} 
\renewcommand{\arraystretch}{1.1} 
\begin{tabularx}{\columnwidth}{@{}ll>{\raggedright\arraybackslash}X@{}} 
\toprule
\textbf{Target STIX Tag} & \textbf{Mapping Strategy} & \textbf{Example Mapped Original Tag(s)*} \\ 
\midrule
\stixtag{Threat-Actor} & Consolidation & \texttt{HackOrg}(D), \texttt{APT}(A), \texttt{THREAT\_ACTOR}(T) \\ 
\midrule
\stixtag{Malware} & Differentiation & \texttt{Malware}(C), \texttt{SamFile}(D), \texttt{MAL}(A) \\
\stixtag{Tool} & Differentiation & \texttt{Tool}(D), \texttt{TOOL}(A), \texttt{ATTACK\_TOOL}(T) \\
\stixtag{Software} & Differentiation & \texttt{System}(C), \texttt{OS}(A) \\ 
\midrule
\stixtag{File} & SCO Mapping & \texttt{FILE}(A), \texttt{SHA2}(A), \texttt{MD5}(A), \texttt{SHA1}(A) \\ 
\midrule
O (Outside) & Exclusion & \texttt{S}(A), \texttt{IDTYL}(A), \texttt{IMPACT}(T) \\ 
\bottomrule
\multicolumn{3}{@{}p{\dimexpr\columnwidth-2\tabcolsep\relax}@{}}{%
  \footnotesize *Sources: (C) CyNER, (D) DNRTI, (A) APTNER, (T) Attacker. Full mapping in repository.%
} \\ 
\end{tabularx}
\vspace{-1.5em} 
\end{table}

\subsection{Schema Mapping and Dataset Construction}
\label{sec:mapping_and_construction}

The core harmonization step involved developing and applying rules to translate original entity tags from each source dataset onto the unified STIX schema. This process addressed the heterogeneity identified previously and leveraged the defined target taxonomy.

The mapping process was guided by the official STIX 2.1 object definitions~\cite{stix_intro} and informed by manual inspection of entity examples within the source datasets to ensure semantic accuracy. We developed explicit translation rules for each source dataset. To represent these mappings consistently and handle tags not corresponding to our target STIX schema, we adopted the standard BIO (Beginning, Inside, Outside) tagging scheme~\cite{ramshaw1995text}. Within this scheme, each token's original tag was mapped either to a corresponding target STIX tag (prefixed with B- or I- to denote entity boundaries) or explicitly designated as 'O' (Outside) if exclusion was warranted.

Key strategies were employed to make these mapping decisions and handle the identified heterogeneity:

\begin{itemize}
    \item \textbf{Consolidation:} Multiple source tags representing the same concept were mapped to a single target STIX tag (e.g., various threat actor tags like \texttt{APT} and \texttt{HackOrg} consolidated to \texttt{Threat-Actor}; diverse vulnerability tags like \texttt{Exp} and \texttt{VULID} unified under \texttt{Vulnerability}).

    \item \textbf{Differentiation:} Related but distinct concepts were mapped to separate STIX objects based on their definitions (e.g., distinguishing between \texttt{Malware}, \texttt{Tool}, and legitimate \texttt{Software}).

    \item \textbf{Mapping to Observables:} Technical indicators were mapped to specific STIX Cyber Observable (SCO) types (e.g., \texttt{IP} $\rightarrow$ \texttt{IPv4-Addr}).

    \item \textbf{Ambiguity Resolution:} Ambiguous source tags were resolved based on STIX definitions and manual review of examples.

    \item \textbf{Exclusion:} Rare, unclear, or irrelevant source tags (based on STIX relevance and frequency) were mapped to the 'O' tag within the BIO scheme, effectively removing them from specific entity status in the harmonized dataset.
\end{itemize}

The defined mapping rules were consistently applied using automated scripts to transform each source dataset's annotations according to the target STIX schema and BIO structure. We then constructed the unified CyberNER corpus by combining this transformed data from all sources. Finally, for usability and compatibility with standard NER tools, the corpus is presented in a CoNLL-like format~\cite{sang2000introduction}, ensuring both clear entity boundaries and structural consistency.

The entire harmonization process is formalized in Algorithm~\ref{alg:harmonization}.

\begin{algorithm}[htbp]
\caption{CyberNER Harmonization and Construction Pipeline}
\label{alg:harmonization}
\begin{algorithmic}[1] 
    \REQUIRE A set of source datasets $D = \{d_1, d_2, ..., d_n\}$
    \ENSURE Harmonized CyberNER corpus $C_{unified}$

    \STATE Analyze schema $S_i$ and entity definitions for each dataset $d_i \in D$.
    \STATE Define target taxonomy $T_{STIX}$ based on STIX 2.1 objects.
    \STATE Initialize an empty mapping rule set $M \leftarrow \emptyset$.
    \FOR{each unique source tag $t_{source} \in \bigcup_{i=1}^{n} S_i$}
        \STATE Map $t_{source}$ to a target tag $t_{target} \in T_{STIX}$ or to 'O' (Outside) based on STIX definitions and manual review of contextual examples.
        \STATE Add rule $(t_{source} \rightarrow t_{target})$ to $M$.
    \ENDFOR
    \STATE Initialize empty corpus $C_{unified} \leftarrow \emptyset$.
    \FOR{each dataset $d_i \in D$}
        \STATE Apply rules in $M$ to transform all token annotations in $d_i$.
        \STATE Convert transformed annotations to a consistent BIO format.
        \STATE Append the processed data from $d_i$ to $C_{unified}$.
    \ENDFOR
    \STATE Perform final quality control and cleaning on $C_{unified}$.
    \STATE \textbf{return} $C_{unified}$
\end{algorithmic}
\end{algorithm}

\section{CyberNER Corpus}
\label{sec:dataset_description}
The harmonization process resulted in the CyberNER corpus, a large-scale, unified resource designed for cybersecurity NER, aligned with the STIX 2.1 standard. This section details its statistical characteristics.

\subsection{Statistics}
\label{sec:statistics}

The final CyberNER corpus represents a substantial aggregation of the source datasets. Overall size and annotation density statistics are summarized in Table~\ref{tab:overall_stats}. The corpus comprises 609,922 tokens organized into 23,477  sentences based on the original dataset structures. Within this, 103,248 tokens are annotated with specific entity tags (excluding 'O'), resulting in an annotation density of 16.93\%, which reflects the typical mixture of specific entities and descriptive text found in cybersecurity documents. The corpus features a case-sensitive vocabulary of 23,432 unique tokens.

\begin{table}[htbp]
\centering
\caption{Overall Statistics of the CyberNER Corpus.}
\label{tab:overall_stats}
\footnotesize 
\begin{tabular}{@{}lr@{}} 
\toprule
Metric                          & Value        \\ \midrule
Total Tokens                    & 609,922      \\
Total Sentences                 & 23,477       \\
Total Annotated Entity Tokens   & 103,248      \\
Annotation Density (\%)         & 16.93\%      \\
Vocabulary Size (Unique Tokens) & 23,432       \\ \bottomrule
\end{tabular}
\vspace{-1em} 
\end{table}

The unified annotation schema, derived from STIX 2.1, encompasses \textbf{21 distinct entity types}. These cover a wide range of cybersecurity concepts, including threat actors, malware, tools, vulnerabilities, attack patterns, indicators, infrastructure, identities, locations, and various cyber observables, providing a comprehensive taxonomy for NER tasks.

\begin{figure}[htbp]
\centering
\includegraphics[width=1.0\columnwidth]{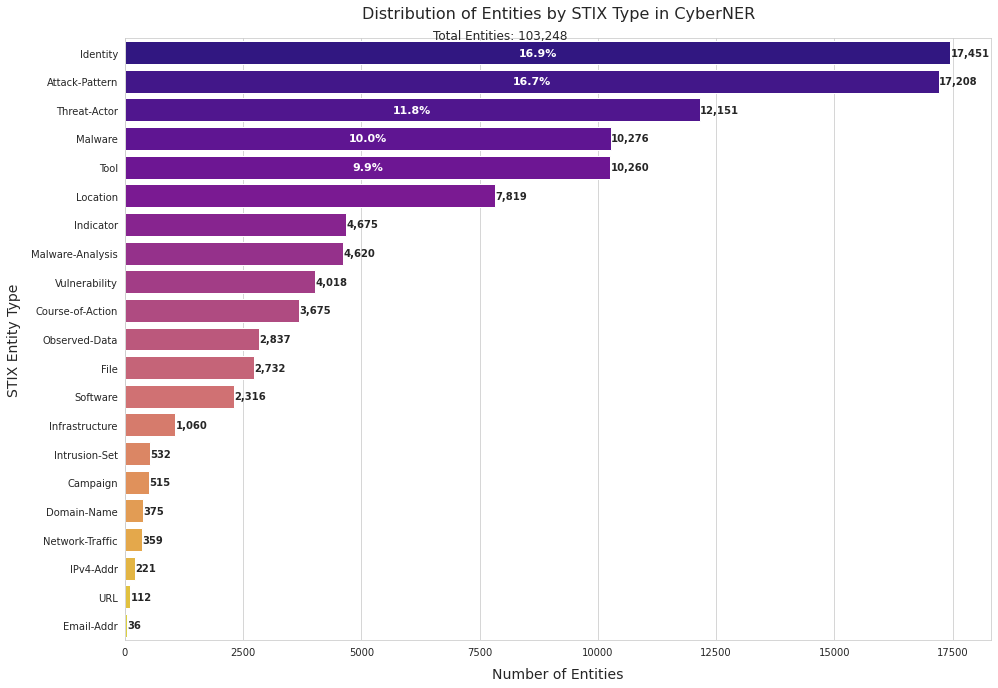} 
\caption{Distribution of Annotated Entities by Unified STIX Type in the CyberNER Corpus.}
\label{fig:entity_dist}
\vspace{-0em}
\end{figure}

A notable characteristic of CyberNER is the distribution of these 21 entity types, illustrated in Figure~\ref{fig:entity_dist}. The distribution shows a significant class imbalance, typical of domain-specific datasets. Core concepts such as \texttt{Identity}, \texttt{Attack-Pattern}, \texttt{Threat-Actor}, \texttt{Malware}, and \texttt{Tool} are highly prevalent, collectively representing over 65\% of all entity annotations. This imbalance presents challenges for models but reflects real-world CTI data.

The composition of CyberNER reflects the diverse origins of its constituent datasets, as shown in Table~\ref{tab:source_contribution}. While APTNER contributes the most tokens (42.7\%), the labeled entities are more evenly sourced from APTNER (34.7\%), DNRTI (30.8\%), and Attacker (27.2\%), with CyNER contributing fewer (7.4\%) due to its sparser schema. This integration yields a corpus with broader conceptual coverage than any individual source. For example, APTNER provides the majority of granular technical observables (\texttt{File}, \texttt{Domain-Name}, hashes mapped to \texttt{File}, etc.), Attacker contributes essential higher-level CTI concepts (\texttt{Campaign}, \texttt{Course-of-Action}, \texttt{Intrusion-Set}), DNRTI is a key source for \texttt{Malware-Analysis} annotations, and CyNER adds coverage for \texttt{Indicator} and \texttt{Software}. This demonstrates the value of harmonization in creating a comprehensive benchmark resource.

\begin{table}[htbp]
\centering
\caption{Contribution of Source Datasets to the CyberNER Corpus.}
\label{tab:source_contribution}
\footnotesize
\begin{tabular}{@{}lrrrr@{}}
\toprule
Source     & Token Count & Token (\%) & Entity Count & Entity (\%) \\ \midrule
CyNER      & 106,991     & 17.54\%    & 7,591        & 7.35\%      \\
DNRTI      & 175,679     & 28.80\%    & 31,799       & 30.80\%     \\
APTNER & 260,290     & 42.68\%    & 35,821       & 34.69\%     \\
Attacker   & 66,962      & 10.98\%    & 28,037       & 27.16\%     \\ \midrule
\textbf{Total} & \textbf{609,922} & \textbf{100.00\%} & \textbf{103,248} & \textbf{100.00\%} \\ \bottomrule
\end{tabular}
\vspace{-1em}
\end{table}

\section{Dataset Benchmarking}
\label{sec:experiments}

To demonstrate the utility of the CyberNER corpus and establish baseline performance benchmarks, we conducted experiments comparing NER models trained on our unified corpus against models trained separately on the individual source datasets.

\subsection{Experimental Setup}

\textbf{Datasets and Training Methodologies:}

We compare two primary training methodologies based on the data used:

\begin{enumerate}
    \item \textbf{Individual Source Training (Baseline):} Models were trained independently for each source dataset (APTNER~\cite{wang2022aptner}, CyNER~\cite{alam2022cyner}, AttackER~\cite{deka2024attacker}, or DNRTI~\cite{wang2020dnrti}), using only data originating from that specific source.

    \item \textbf{Unified CyberNER Training:} A single model was trained using the complete, harmonized CyberNER corpus, which integrates data from all source datasets.
\end{enumerate}

\textbf{Model Architecture:}
We adopted a standard and effective architecture for sequence labeling, combining a pre-trained transformer model with a Conditional Random Field (CRF) layer~\cite{chen2023enhancing}. The transformer encoder generates contextualized token representations, which are then fed into a linear layer followed by the CRF~\cite{lafferty2001conditional}. The CRF layer explicitly models dependencies between adjacent output tags, learning transition constraints to improve the overall coherence and accuracy of the predicted entity sequences. Different pre-trained transformer backbones were evaluated in our benchmark, as detailed in Section~\ref{sec:results_analysis}. 

\textbf{Hyperparameters.} We used the AdamW optimizer with differential learning rates (\textit{encoder}: $5\mathrm{e}{-5}$, \textit{CRF/FC}: $8\mathrm{e}{-5}$) and corresponding weight decay (\textit{encoder}: $1\mathrm{e}{-5}$, \textit{CRF/FC}: $5\mathrm{e}{-6}$, excluding bias and normalization parameters). A linear warmup (10\% of training steps) preceded a linear learning rate decay~\cite{devlin2019bert}. Key settings included a maximum sequence length of 256, batch size of 16, and gradient norm clipping at 1.0. Models were trained for up to 50 epochs with early stopping (patience of 5 epochs based on validation F1), and the best validation checkpoint was used for final testing.

\textbf{Evaluation Metrics:}
Models were evaluated using standard entity-level Precision (P), Recall (R), and micro-averaged F1-score (F1). The overall micro-F1 served as the primary comparison metric, while Precision and Recall provided further diagnostic information. To provide additional insights beyond token-level accuracy, particularly for models trained on CyberNER, we report sentence-level coverage metrics including the average predicted entities per sentence (Avg Ent/Sent), micro-averaged Sentence Entity Recall (Sent Recall (Micro \%)), and macro-averaged Sentence Entity Recall (Sent Recall (Macro \%)).

\subsection{Baseline Performance on Source Datasets}
\label{sec:results_analysis}

\begin{figure}[htbp]
\centering
\begin{adjustbox}{width=\columnwidth, center} 
\includegraphics{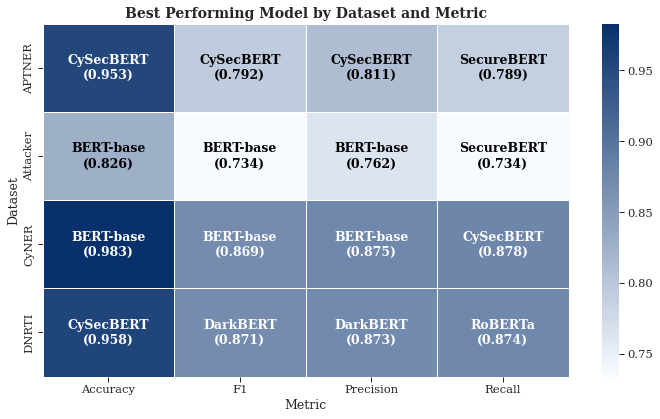} 
\end{adjustbox}
\caption{Best Performing Model and Score for Each Source Dataset across Different Evaluation Metrics (Accuracy, F1, Precision, Recall). Models were evaluated independently on original, non-harmonized data.}
\label{fig:best_model_heatmap}
\vspace{0em} 
\end{figure}
To establish performance baselines prior to harmonization, we evaluated five pre-trained transformer models (BERT-base-cased~\cite{devlin2019bert}, CySecBERT~\cite{fiorini2023cysecbert}, DarkBERT~\cite{jin2023darkbert}, RoBERTa-base~\cite{liu2019roberta}, and SecureBERT~\cite{aghaei2022securebert}) adapted for NER via a CRF layer. Each model was fine-tuned and tested independently on the original schemas and data splits of the four source datasets: APTNER, Attacker, CyNER, and DNRTI.

Figure~\ref{fig:best_model_heatmap} visualizes the top-performing model for each source dataset across key metrics (Accuracy, F1, Precision, Recall). The results highlight considerable variability. Firstly, baseline performance differs significantly across datasets, with models achieving higher scores on CyNER and DNRTI compared to APTNER and Attacker, likely due to variations in schema complexity, annotation density, and text domain. Secondly, and crucially, no single pre-trained model consistently dominates across all datasets or metrics. For instance, while CySecBERT achieves the best F1-score on APTNER (0.792), BERT-base leads on Attacker (0.734) and CyNER (0.869), and DarkBERT leads on DNRTI (0.871). Furthermore, the optimal model often changes depending on the chosen metric even within the same dataset (e.g., on DNRTI, DarkBERT leads F1/Precision, while RoBERTa leads Recall). This inconsistency, where no single model architecture excels across all domains, strongly implies that models specialized on one schema fail to generalize to others, further motivating the need for a unified training approach.

\subsection{Baseline: The Failure of Naive Concatenation}
To empirically validate the central premise of our work, we first established a baseline by training models on a \textit{naively concatenated} version of the four source datasets. In this setup, a single model was tasked with learning from the original, unharmonized set of over 50 unique entity tags (e.g., `APT`, `HackOrg`, `MAL`, `SamFile`, etc.) simultaneously. This experiment simulates the outcome of simply merging available data without a principled harmonization strategy.

The results, summarized in Table~\ref{tab:naive_baseline_results}, highlight the severe limitations of this approach. The overall F1-scores are low, with the best model, RoBERTa-base, achieving only \textbf{0.569 F1}. Furthermore, performance is highly volatile across the different data sources. For instance, while models perform adequately on the simpler CyNER schema, their F1-scores plummet to as low as 0.350 on the more complex APTNER subset. This demonstrates that the model is unable to form a coherent, generalizable understanding of cybersecurity entities from the noisy and contradictory label space, confirming that naive concatenation is an inadequate solution.

\begin{table}[htbp]
\centering
\caption{Performance of Naive Concatenation Baseline (Training on Unharmonized Labels). Models struggle with low overall scores and high variance across source-specific test subsets.}
\label{tab:naive_baseline_results}
\footnotesize
\begin{adjustbox}{width=\columnwidth, center}
\begin{tabular}{@{}lccccc@{}}
\toprule
\textbf{Model} & \textbf{Overall F1} & \textbf{APTNER} & \textbf{CyNER} & \textbf{Attacker} & \textbf{DNRTI} \\ \midrule
BERT-base      & 0.558 & 0.381 & 0.767 & 0.628 & 0.634 \\
RoBERTa-base   & \textbf{0.569} & 0.448 & \textbf{0.784} & 0.623 & 0.602 \\
SecureBERT     & 0.559 & 0.350 & 0.740 & 0.611 & \textbf{0.691} \\
CySecBERT      & 0.551 & \textbf{0.539} & 0.774 & 0.595 & 0.477 \\
DarkBERT       & 0.556 & 0.428 & 0.741 & \textbf{0.614} & 0.601 \\ \bottomrule
\end{tabular}
\end{adjustbox}
\vspace{-1em}
\end{table}

\subsection{Performance Evaluation on CyberNER}
\label{sec:cyberner_evaluation}

Having established baselines on the original datasets, we now evaluate the performance of models trained on the unified CyberNER corpus. This assesses the models' ability to handle the increased complexity and diversity introduced by harmonization and the STIX schema.

\begin{figure}[htbp]
\centering
\includegraphics[width=0.9\columnwidth]{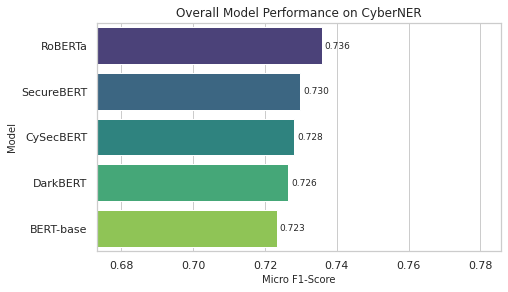} 
\caption{Overall Performance (Micro F1-Score) of models trained and evaluated on the unified CyberNER test set.}
\label{fig:overall_cyberner_f1}
\vspace{-0em} 
\end{figure}

Figure~\ref{fig:overall_cyberner_f1} presents the overall micro F1-scores achieved by the five evaluated models on the held-out CyberNER test set. RoBERTa demonstrates the highest performance (0.736 F1), closely followed by SecureBERT (0.730), CySecBERT (0.728), DarkBERT (0.726), and BERT-base (0.723). Notably, these overall scores on the unified task are lower than the peak scores achieved by the same models when trained and tested on individual, simpler source datasets (e.g., baseline F1 scores often exceeded 0.85 on CyNER or DNRTI, see Section~\ref{sec:results_analysis}). This performance difference is expected, reflecting the significantly more challenging nature of the unified task, which demands recognition of 21 distinct STIX entity types derived from multiple text domains, compared to the narrower scope of the original datasets.

\begin{figure}[htbp]
\centering
\includegraphics[width=\columnwidth]{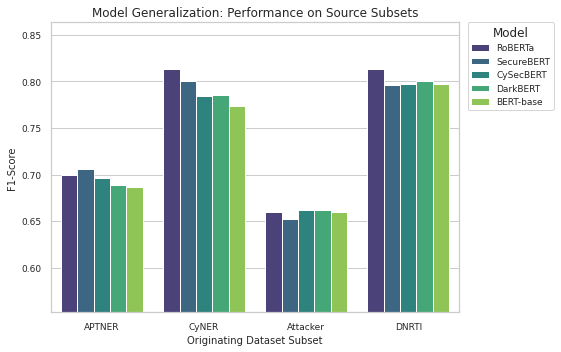}
\caption{Source-Specific Performance (F1-Score) on test subsets for models trained on the unified CyberNER corpus.}
\label{fig:source_specific_f1}
\vspace{0em} 
\end{figure}

To analyze generalization, Figure~\ref{fig:source_specific_f1} shows the performance of these unified models when evaluated on test subsets grouped by their original source. RoBERTa again shows strong average performance across sources (0.746 average F1), but all models exhibit variability, generally performing better on subsets originating from CyNER and DNRTI compared to the Attacker subset. Critically, comparing these scores to the baseline results reveals the trade-offs of unification. For instance, RoBERTa trained on CyberNER achieved an F1 of 0.699 on the APTNER subset, whereas RoBERTa trained solely on APTNER reached 0.781 (Section~\ref{sec:results_analysis}). Similarly, DarkBERT achieved 0.800 F1 on the DNRTI subset when trained on CyberNER, compared to its baseline of 0.871 when trained only on DNRTI. This consistent pattern underscores that while unified models gain breadth and generalizability, they may sacrifice some specialized performance compared to models trained exclusively on a single source domain's specific schema and style. This highlights CyberNER's value in training broadly applicable models, while acknowledging that peak performance on a narrow task might still require domain-specific training.

Finally, we examined sentence-level coverage metrics (Table~\ref{tab:coverage_metrics}) to assess information capture beyond F1. Although RoBERTa led in F1, BERT-base predicted the most entities per sentence on average (2.67), with DarkBERT also showing relatively high output density (2.53). Sentence recall scores remained relatively consistent across all models (~72\% Micro, ~72\% Macro), suggesting similar overall entity finding capabilities at the sentence level among the different architectures. These coverage metrics offer a complementary view, indicating that optimizing solely for F1 might not maximize the density of extracted information, a potentially important factor for CTI analysts seeking comprehensive data extraction.

\begin{table}[htbp]
\centering
\caption{Key evaluation metrics, including sentence-level coverage, for models trained on CyberNER.}
\label{tab:coverage_metrics}
\footnotesize
\setlength{\tabcolsep}{4pt} 
\renewcommand{\arraystretch}{1.1} 
\begin{tabular}{@{}lcccc@{}}
\toprule
\textbf{Model} & \textbf{F1} & \makecell[c]{\textbf{Avg} \\ \textbf{Ent/Sent}} & \makecell[c]{\textbf{Sent Recall} \\ \textbf{(Micro \%)}} & \makecell[c]{\textbf{Sent Recall} \\ \textbf{(Macro \%)}} \\
\midrule
RoBERTa     & 0.736 & 2.44 & 72.1\% & 72.5\% \\
SecureBERT  & 0.730 & 2.42 & 71.8\% & 72.4\% \\
CySecBERT   & 0.728 & 2.47 & 71.8\% & 71.7\% \\
DarkBERT    & 0.726 & 2.53 & 71.6\% & 72.0\% \\
BERT-base   & 0.723 & 2.67 & 72.7\% & 72.0\% \\
\bottomrule
\end{tabular}
\vspace{-1em}
\end{table}

Overall, these benchmark results validate CyberNER as a challenging but valuable resource. Models trained on it demonstrate reasonable generalization, Though with some performance trade-off compared to highly specialized single-source models. The unified corpus enables the development and standardized comparison of NER systems with broad, STIX-aligned entity coverage, essential for advancing automated CTI analysis.

\section{Discussion}
\label{sec:discussion}

Our experimental results provide compelling evidence that a principled, ontology-driven harmonization is not merely a beneficial step but a necessary one for building robust, general-purpose NER models in cybersecurity. The stark contrast between the naive concatenation baseline (0.569 F1) and the performance on CyberNER (0.736 F1) highlights the severe penalty of schema heterogeneity. By mapping disparate source labels to the STIX 2.1 standard, we provided a coherent supervisory signal that enabled models to achieve a relative performance gain of approximately 30\%. This demonstrates that the primary challenge is not a lack of data, but a lack of semantic consistency across available datasets. The STIX 2.1 taxonomy proved to be a highly effective framework for this task, offering a standardized vocabulary that enhances both model performance and the interoperability of NER outputs with other CTI platforms.

It is important to acknowledge the trade-off between generalization and specialization. While models trained on the unified CyberNER corpus significantly outperform the naive baseline and show strong generalization across sources, a model hyper-specialized on a single dataset's original schema (e.g., trained and tested only on CyNER) may still achieve a higher score on that specific, narrow task. However, such a specialist model lacks the broad entity coverage and interoperability essential for real-world CTI applications. CyberNER successfully bridges this gap, enabling the development of models that are broadly effective across diverse textual domains, a feat unattainable through simple data aggregation.

Despite these benefits, the CyberNER corpus inherits certain limitations. Its quality is fundamentally dependent on the underlying source datasets; any annotation errors or biases present in CyNER, DNRTI, APTNER, or Attacker are potentially propagated into the unified corpus. Furthermore, while comprehensive, the STIX standard sometimes required the consolidation of fine-grained distinctions from a source schema into a broader STIX category. The current scope of CyberNER is also limited to the four integrated datasets; incorporating additional CTI NER resources could further enhance coverage but requires extending the harmonization effort. Furthermore, while our baseline results imply poor generalization of specialized models, a comprehensive cross-dataset evaluation remains an important area for future work to quantitatively measure this effect.

\textbf{Future Work:} 
Future directions for this research are threefold, focusing on the corpus, methodology, and task extension. First, the CyberNER corpus itself can be expanded by incorporating additional public datasets, such as STUCCO, as well as proprietary industry data to further enhance its diversity and coverage. Refining the STIX mapping based on community feedback and exploring cross-lingual harmonization also remain important avenues.

Second, future methodological work can explore the performance trade-off between generalization and specialization. This includes conducting a detailed ablation study to disentangle the effects of schema harmonization from increased task complexity, as well as investigating advanced training strategies like multi-task learning to potentially mitigate this trade-off.

Finally, a significant next step is to extend the annotation beyond entity recognition to include relation extraction, aligned with STIX Relationship Objects (SROs). This would enable the automated construction of rich cybersecurity knowledge graphs directly from text, unlocking deeper CTI analysis.

\section{Conclusion}
\label{sec:conclusion}

This paper addressed the critical challenge of schema heterogeneity that has hindered progress in cybersecurity Named Entity Recognition. We demonstrated that naively concatenating existing datasets with their disparate, original labels results in poor model performance and a failure to generalize across different data sources. 

To overcome this, we introduced \textbf{CyberNER}, a large-scale corpus created by systematically harmonizing four prominent NER datasets (CyNER, DNRTI, APTNER, and Attacker) onto a unified taxonomy based on the STIX 2.1 standard. Our empirical evaluation showed that this principled harmonization is highly effective. Models trained on CyberNER achieved a substantial performance gain, with a relative F1-score improvement of approximately 30\% over the naive concatenation baseline.

By publicly releasing CyberNER, we provide the community with a valuable resource for training more robust, broadly applicable cybersecurity NER models. More importantly, it serves as a standardized benchmark that enables rigorous and meaningful comparisons for future research, paving the way for more advanced and interoperable automated CTI systems. Future work will focus on expanding CyberNER with additional datasets and exploring relation extraction capabilities.

\section*{Acknowledgment}
The authors would like to thank Deloitte Morocco Cyber Center and Deloitte Conseil for their support of this research. 

\bibliographystyle{IEEEtran}
\bibliography{references}

\end{document}